\documentstyle[preprint,aps,epsfig]{revtex}
\def\pmb#1{\setbox0=\hbox{#1}
  \kern-.015em\copy0\kern-\wd0
  \kern.03em\copy0\kern-\wd0
  \kern-.015em\raise.0233em\box0}

\def\bm#1{\pmb{${#1}$}}
\begin{document}
\draft
\title{PRODUCTION OF LIKE SIGN DI-LEPTONS  IN \bm{p-p}\  COLLISIONS
THROUGH COMPOSITE MAJORANA NEUTRINOS.
}
%
\author{O.~Panella\footnote{Author to whom correspondence should be 
addressed. \\ Electronic address:
{\tt Orlando.Panella@PG.infn.it}}$^{(a)}$, C.~Carimalo$^{(b)}$ and 
Y.~N.~Srivastava$^{(a,c)}$
}
\author{}
\address{
$^{a)}$Istituto Nazionale di Fisica Nucleare, Sezione di Perugia,\\
 Via A.~Pascoli, I-06123 Perugia, Italy\\
}
\author{}
\address{
$^{b)}$Laboratoire de Physique Nucl\'eaire et de Hautes Energies, IN2P3-CNRS\\ 
Universit\'es Paris VI/VII,\\
4 place Jussieu, F-75252, Paris Cedex 05, France\\
}
\author{}
\address{
$^{c)}$ Northeastern University, Physics Department, Boston Ma, 02115\\
}
\date{March 4, 1999}
\maketitle
\newpage
\begin{abstract}
The production of Like-Sign-Di-leptons 
(LSD), in the high energy lepton number violating ($\Delta L = +2$) 
reaction, $\, p p \to 2\, \hbox{jets} + \ell^+\ell^+$, ($\ell=e,\mu,\tau$), 
of interest for 
the experiments to be performed at the forthcoming Large Hadron Collider (LHC),
is investigated in detail, taking up
a composite model scenario in which the exchanged virtual {\it composite} 
neutrino  is assumed to be a Majorana particle 
that couples to the light leptons via the $SU(2)\times U(1)$ gauge 
bosons through a magnetic type coupling ($\sigma_{\mu\nu}$). 
An helicity projection method is used to evaluate exactly the tree-level
amplitudes of the contributing parton subprocesses ($2\to 4$), which allows
to take into account  all exchange diagrams and occurring 
interferences. Numerical estimates of the corresponding signal
cross-section that implement   kinematical cuts needed to 
suppress the Standard Model background, are presented which show that
in some regions of the parameter space the total number of LSD events 
is well above the background. 
Assuming non-observation of the LSD signal it is found that LHC would
exclude a composite Majorana neutrino up to 850 GeV (if one requires
10 events for discovery).
The sensitivity of LHC experiments to the parameter space is then compared
to that  of the
next generation of neutrinoless double beta decay ($\beta\beta_{0\nu}$)
experiment,  GENIUS, and it is shown  that they will provide 
constraints of the same order of magnitude and  will  play a 
complementary role.
\end{abstract}


\section{Introduction}
Since the discovery of the $Z^{0}$ and $W^{\pm}$ gauge bosons~\cite{rubbia}  
the standard model (SM) of electroweak interactions~\cite{sm}  based on the 
SU(2)$\times$U(1) gauge group has scored an impressive record of experimental
checks. However some unexplained facts of the model like the mass hierarchy, 
the proliferation of elementary particles,
and the total number of free parameters have lead to believe  that it is
only a low energy manifestation of a yet unknown underlying 
fundamental theory, which 
would be free of the above theoretical difficulties.
Therefore despite the enormous experimental success of the SM many alternative
theories have been developed such as Left-Right symmetric models, 
composite models, super-symmetry, string theory, grand unified models.
The investigation of  effects predicted by the new theories that are absent 
in the standard theory is therefore very important since, were these effects 
to be experimentally observed they would signal {\it new physics} 
unaccounted for by the SM.
It is in this direction that a great portion of recent theoretical 
and experimental studies have been concentrated~\cite{HM}, and this is indeed 
the spirit of this work which deals with lepton number violating processes.

The conservation of the total lepton number ($L$) is one of the symmetries
of the SM experimentally observed to hold true until now.
In the SM
with massless Dirac neutrinos 
processes with $\Delta L \ne 0$ are not possible. 
Violation  of this symmetry is generally related to
the existence of massive Majorana particles and many  
extensions of the  SM contain
$L$-violating interactions involving Majorana neutrinos.
Left-Right symmetric models for example
contain right-handed Majorana neutrinos, with a mass that could be in the 
TeV range, and  coupled to the light leptons via the right-handed 
gauge bosons ($W_R, Z_R$)~\cite{pati}.
Superstring generated $E_6$ models also have neutral 
Majorana leptons~\cite{hewett}. 
Finally ref.~\cite{masiero} provides an example of a composite model with
Majorana neutrals.

The effect which seems most promising with respect to showing  
violations of the lepton number
is the neutrinoless double beta decay ($\beta\beta_{0\nu}$), 
a second order process where, in a nucleus, two protons 
(neutrons) undergo simultaneously a weak beta decay emitting two 
positrons (electrons) while the two neutrinos annihilate 
into the vacuum~\cite{HM}:
\begin{equation}
A(Z+2) \to A(Z) + e^+ e^+   \quad \quad \Delta L = + 2
\label{0nubb}
\end{equation}
This process is
only possible if the  neutrino is a massive Majorana
particle, and thus it is impossible within the SM. 
Experiments that search for such rare decay have since long being 
performed but always with negative results~\cite{bbexp}. 
Currently the Heidelberg-Moscow $\beta \beta$ experiment 
at the Gran-Sasso laboratory in Italy provides the best experimental 
lower bound on the half-life of the process~\cite{klapring}:
\begin{eqnarray}
^{76}\hbox{Ge} & \to &  ^{76}\hbox{Se} + 2\, e^- \nonumber \\ 
T^{\beta\beta_{0\nu}}_{1/2} & > &  1.2\times 
10^{25}\hbox{yr} .
\label{bbbound}
\end{eqnarray}
The proposed GENIUS double beta experiment (see section VI), 
now under development, will 
either increase the lower bound on the half-life by two or three orders
of magnitude or observe the decay.
From the theoretical point of view,
the strong bound on the half-life in Eq.~(\ref{bbbound}) has 
been turned into a powerful tool
to impose constraints on models of new physics which predict a non 
zero amplitude for the $\beta\beta_{0\nu}$ decay~\cite{trento}.
Studies in this direction include:
an investigation of new super-symmetric contributions
from R-parity violating MSSM~\cite{hirsch1} which shows how 
constraints on parameters of the model from  non-observation
of $\beta\beta_{0\nu}$ are stronger than those available from
accelerator experiments;
a  detailed analysis of the contribution to $\beta\beta_{0\nu}$ 
from left-right symmetric models\cite{hirsch2};  
a study of the effective low energy 
charged current lepton-quark interactions due to the exchange
of heavy leptoquarks~\cite{hirsch3}. 

The present authors have, in a series of recent papers~\cite{op1,op2,op3},
investigated the contribution, to the neutrinoless double beta decay, of a 
heavy Majorana neutrino, arising from a composite model scenario in which
the excited partner of the neutrino (the excited neutrino, $\nu^\star$)
is a assumed to be a Majorana particle. 
This study revealed that $\beta\beta_{0\nu}$ constraints are competitive, 
and in some regions of the parameter space, even 
more restrictive than those derived from high-energy direct
search of excited particles~\cite{op3,opring}.
This result  led to consider the potential of the experiments to
be performed at the forthcoming Large Hadron Collider (LHC) at CERN,
with respect to the possibility of observing the production of Like Sign
Di-leptons, $ \ell^+\ell^+$ or $ \ell^-\ell^-$, $\ell=e,\mu,\tau$,
(hereafter denoted LSD) 
in proton-proton collisions with an energy of $14$ TeV 
in the center of mass frame:
\begin{equation}
p p \to 2\hbox{jets} + \hbox{LSD} , \qquad \Delta L = +2.
\end{equation}
In hadronic collisions LSD can be produced in quark-quark 
(antiquark-antiquark) scattering, through the elementary sub-process
$W^+ W^+ \to \ell^+\ell^+$ (virtual $W$-boson fusion) as depicted in 
Fig.~\ref{fig1} where the dashed blob represents all contributing diagrams 
within a given model.
As regards this mechanism of LSD production one can say that
it is the high-energy analog of the neutrinoless double beta 
decay which indeed proceeds through the same Feynman diagrams 
(see for example ref.~\cite{op3}). Fig.~\ref{fig2}a shows explicitly the 
Feynman diagram for the production of LSD through the 
exchange of a heavy Majorana neutrino, (basic mechanism).  
In the case of quark-antiquark scattering in addition 
to the $W$ fusion another mechanism must be considered that leads 
to LSD production: 
the direct production of a heavy Majorana neutrino 
via quark-antiquark annihilation,  $ q \bar{q}' \to \ell^+ N $, with the  
subsequent decay of the heavy neutrino $N \to \ell^+ q \bar{q}'$ 
(annihilation mechanism). This is depicted in Fig.~\ref{fig2}b.  

Production of LSD has been considered in the past by several authors and
within the context of different models. In the Left-Right symmetric
model, Keung and Senjanovi\'c~\cite{keung} already in 1983 realized that 
the associate production of LSD with two hadronic jets would signal the 
annihilation of quark-antiquark pairs into the right-handed gauge 
boson of the model ($W_R$). Estimates were given for $p p$ collisions
at $\sqrt{s}=800$ GeV. The study  of this model was later taken up to
higher energies (SSC and LHC) by Datta, Guchait and Roy in \cite{datta} 
where the authors indicated how to effectively reduce 
the SM background.
Dicus, Karatas and Roy~\cite{dicus}  have studied LSD production 
at high-energy
hadron colliders through the exchange of heavy right-handed Majorana 
neutrinos, without commitment to a specific model (beyond the SM). They used
a $\gamma_{\mu} $-type coupling and found the LSD signal  detectable 
at the SSC while at the LHC the SM background would probably preclude
detection.
Two of the present authors~\cite{opthesis} provided a rough estimate of  
the signal cross sections for $ p p \to 2\hbox{jets + LSD} $ at LHC 
within the context of 
composite models (exchange of a heavy composite Majorana neutrino with a
$\sigma_{\mu\nu}$-type coupling)
using an equivalent W-boson approximation~\cite{dawson} (similar 
to the Weisz\"acker Williams approximation for the photon field) 
and integrating over the complete 
phase-space of the 
sub-process $W^+ W^+ \to \ell^+\ell^+$. The result was that the signal
could be observable at the LHC. 

One remark should be made at this point that applies to all works just cited
that have investigated LSD production in $pp$ collisions. 
None of them deals, {\it at the same time}, with the two mechanisms of LSD 
production, i.e. $W^+W^+$-fusion and $q\bar{q}'$-annihilation.
Indeed when dealing with $q\bar{q}'$ scattering both mechanisms must be 
considered and the corresponding amplitudes should be added coherently.
In order to do so one needs a way of efficiently
computing the amplitudes.
In this paper it is  done precisely so, 
calculating  analytically the helicity amplitudes of the occurring  
tree-level diagrams and  accounting  thus for the interference term between
the   $W^+W^+$-fusion and the $q\bar{q}'$-annihilation.

Thus the goal of this paper is twofold: (i) to address the sensitivity
of LHC experiments with respect to the parameters of the composite model 
effective Lagrangian and compare this to that of the next generation of
double beta decay experiments now under development (GENIUS); (ii) 
to present a calculation of LSD production in $pp$ collisions (via 
the exchange of a heavy composite Majorana neutrino) which goes
beyond the approximations of ~\cite{opthesis} and which, in the case of
$q\bar{q}'$-annihilation, includes coherently
the two competing mechanisms.

The rest of the paper is organised as follows: in section II  the reader 
is briefly
reminded  of the effective Lagrangian describing the coupling of 
the excited neutrino with the  electron and  
a comparison between recent bounds on the parameters from the low energy 
$\beta\beta_{0\nu}$ experiment by the Heidelberg-Moscow Collaboration
and  those from high energy experiments  
performed by the DELPHI Collaboration at the
Large Electron Positron (LEP) Collider is presented; 
in section III the amplitudes 
of the L-violating parton sub-processes
are presented; section IV contains:  ($i$) a description 
of the kinematical cuts applied with a short discussion 
of the background; ($ii$)  our numerical results for the 
signal cross-sections; 
in section V  the sensitivity to the
parameter space of LHC is compared to that of GENIUS; 
finally, section VI contains the conclusions.

\section{Compositeness and existing \bm{\beta\beta_{0\nu}}\  constraints.}

It is well known that one possible scenario of physics beyond the SM
is one in which quarks and leptons are not elementary particles
but posses an internal structure, i.e. they are bound states of, yet
unknown, new constituents, generally referred to as {\it preons}, bound 
together by a new dynamical interaction. Theories
that follow this path are called {\it composite models} 
and although many have been proposed~\cite{preons} none has emerged as 
a new dynamically consistent theory.
However there are some model independent consequences of the idea
of compositeness which can be addressed without commitment to any 
specific model. These are: (i)  contact interactions between ordinary 
fermions; (ii) the existence of excited partners for quarks and leptons 
with masses of the order of the compositeness scale, 
$\Lambda_{\hbox{c}}$.
Phenomenologically these ideas have been studied via effective 
interactions~\cite{op2,pdg}. In particular  in this work, the case 
of excited neutrinos ($N$), is taken up  
and only the relevant coupling with the light electron are reviewed.
Effective couplings between the heavy and light leptons (or quarks) have been
proposed, using weak iso-spin 
($I_W$) and hyper-charge ($Y$) conservation~\cite{cab}.
Assuming that such states are grouped in $SU(2) \times U(1)$ 
multiplets, since  light  fermions have $I_W=0,1/2$
and electroweak  gauge bosons have $I_W=0,1$,
only multiplets with $I_W \leq 3/2$ can be excited in 
the lowest order perturbation theory.   
Also, since none of the gauge fields carry hyper-charge, a
given excited multiplet can couple only to a light multiplet with 
the same $Y$. 
The transition coupling of heavy-to-light fermions
is assumed to be of the magnetic moment type respect to any 
electroweak gauge bosons~\cite{cab}.
Restrict here to the first family and consider 
spin-$1/2$ excited states grouped in multiplets with 
$I_W=1/2 $ and $ Y=-1$ (the so called homodoublet model~\cite{pdg}), 
\begin{equation}
L= {N \choose E} 
\end{equation} 
which can couple to the light  left-handed multiplet
\begin{equation}
\ell_L = {\nu_L \choose e_L} ={{1-\gamma_5} \over 2}
{\nu \choose e}
\end{equation}
through the gauge fields ${\vec W}^{\mu}\, \hbox{and} B^{\mu}$.
The relevant interaction is written\cite{cab} 
in terms of two {\it new} independent coupling constants $f$ and $f'$:
\begin{eqnarray}
{\cal L}_{int}& = & \frac{gf}{\Lambda_{\hbox{c}}} \bar{L}\sigma_{\mu\nu}
\frac{\vec\tau}{2} l_L \cdot \partial^\nu\vec{W}^\mu \cr
& & \phantom{xxxxx}
+\frac{g'f'}{\Lambda_{\hbox{c}}}
\biggl(-\frac{1}{2} \bar{L}\sigma_{\mu\nu}
l_L \biggr)\cdot \partial^\nu B^\mu + \hbox{h.c.}
\end{eqnarray}
where ${\vec \tau}$ are the Pauli $SU(2)$
matrices, $g$ and $g'$ are the usual $SU(2)$ and $U(1)$ gauge coupling
constants, and the factor of $-1/2$ in the second term is the
hyper-charge of the $U(1)$ current. 
This effective Lagrangian is  widely used in the literature
to predict production cross sections and decay rates of the excited
particles~\cite{pdg,ruj,baur}.
In terms of the physical gauge fields the interaction Lagrangian 
describing the  coupling of
the heavy excited neutrino with the light electron is therefore:
\begin{equation}
\label{leff}
{\cal L}_{eff} =  \bigl({ g f \over {\sqrt 2} \Lambda_{\hbox{c}}} \bigr)
\Bigl\{ 
\Bigl( {\overline N}
\sigma^{\mu \nu} {1-\gamma_5 \over 2} \, e \Bigr)
\, \partial_{\nu} W_{\mu}^+\Bigl\}\, + \, \hbox{H.c.} 
\end{equation}
In the 
analysis carried out in~\cite{op1,op2} it was assumed that the excited  
neutrino is a Majorana particle with mass $M_N$, expected to be of the order
of the compositeness scale $\Lambda_{\hbox{c}}$, which would then contribute 
to the neutrinoless double beta decay. The following result
for the half-life of the $\beta\beta_{0\nu}$ was found~\cite{op2}:
\begin{equation}
\label{t1/2}
T_{1/2}^{-1} = \left(\frac{f}{\Lambda_{\hbox{c}}}\right)^4 \frac{m_A^8}{M_N^2}
\, |{\cal M}_{FI} |^2 \, \frac{G_{01}}{m_e^2},
\end{equation}
where $m_A = 0.85$ GeV is a parameter entering the nuclear form factors, 
${\cal M}_{FI}= -5.45 \times 10^{-2}$ 
is a nuclear matrix element, $m_e $ is the electron mass
and $G_{01}=6.4 \times 10^{-15}$ yr$^{-1}$ is a phase space integral. 
Combining this result with the non-observation of the decay
($T_{1/2} > T_{1/2}^{\,\, \text{lower bound}}$) one obtains a 
constraint on the parameters of the model:
\begin{equation}
\label{constraint}
\left|\frac{f}{\Lambda_{\hbox{c}}}\right| < M_N^{1/2}
\left(\frac{m_e^2}{m_A^8}\right)^{1/4}
\frac{\left[G_{01}\, T_{1/2}^{\, \, \text{lower bound}}
\, 
\right]^{-1/4}}
{ |{\cal M}_{FI} |^{1/2} }.
\end{equation}
Using the current experimental lower bound on the half-life of the
$^{76}\hbox{Ge} $ decay
provided by the Heidelberg-Moscow $\beta\beta$ experiment, the 
following  
constraint  on the parameters $ f,  \Lambda_{\hbox{c}}, M_N $
appearing in Eq.~(\ref{leff}) is 
deduced~\footnote{This is an updated constraint respect to that 
of ref.~\protect\cite{op1} where a previous value of the half-life 
was used.}:
\begin{equation}
\label{constraint_num}
|f| \le 8.03 
\frac{\Lambda_{\hbox{c}}}{1\, \hbox{TeV}}\biggl(\frac{M_N}{1\, 
\hbox{TeV}}\biggr)^{1/2}.
\end{equation}
This double beta bound on 
compositeness can be compared with 
bounds on the same parameters from high energy experiments
performed at the Large Electron Positron (LEP) collider, phase II.
The DELPHI Collaboration has reported~\cite{delphi} on a search
for excited leptons in $ e^+ e^- $ collisions at $\sqrt{s} = 183$ GeV, 
where both the single and double production mode were studied. It should be 
emphasized that the analysis in \cite{delphi}
was carried out using the same effective Lagrangian that was 
considered in \cite{op1,op2}, c.f. Eq.(\ref{leff}), 
so that it makes sense to compare the corresponding bounds.
In Fig.~\ref{delphi_com} the bound of Eq.~\ref{constraint_num} is plotted
against the exclusion curve of the DELPHI Collaboration~\cite{delphi}, and
one can see that for masses above $\approx 110 $ GeV the double beta bound
is more constraining, i.e. it excludes a portion of parameter phase space
still allowed by the DELPHI exclusion 
plot\footnote{It should be noted that also the ALEPH Collaboration 
has recently published results of a search for compositeness 
at LEP I. In ref.~\cite{aleph} bounds on the 
compositeness scale, in particular regarding  
the same excited neutrino couplings discussed  here, are reported. 
Choosing $f=f'=1$  a  neutrino mass dependent  
lower bound on $\Lambda_{\hbox{c}}$ is found which is about 
$16$ TeV at $M_N={\cal O}(10\hbox{ GeV})$ 
while it drops down to $4$ TeV at the 
maximum value of $M_N$ explored of $80$ GeV. This result is not directly
comparable to Eq.~(\ref{constraint_num}) 
since this was derived within the hypotesis $M_N >> M_W $~\cite{op3}. 
Assuming $ |f| =1$, Eq.~(\ref{constraint_num}) gives :  $\Lambda_{\hbox{c}}\ge
0.12 \hbox{ TeV} $ at $M_N = 1$ TeV.}.
This result prompted the present authors to study the potential of the LHC with
respect to the same type of lepton number violating processes, with an emphasis
on comparing its sensitivity with that of the next generation of double
beta decay experiments.
The following section deals  with
the calculation of the lepton number violating processes 
in $pp$ collisions described by the diagrams of Figures~1 and 2.  
They have been carried out 
with a choice of the parameters that satisfies the bounds from 
$\beta\beta_{0\nu}$ just discussed.

\section{Amplitudes of \bm{L}-violating Parton sub-processes}
In the following  the helicity amplitudes for parton 
sub-processes that contribute to production of LSD via the exchange
(or production) of a heavy Majorana composite neutrino are presented.
The effective interaction used is that of Eq.~(\ref{leff}). 
Considering for the moment only the first fami\-ly, 
three 
different types of processes should be distinguished:
\begin{eqnarray}
&(i)&\,
u u     \to d d     + \ell^+\ell^+, \nonumber \\ 
&(ii)&\,  u \bar{d} \to d \bar{u} + \ell^+\ell^+, \nonumber\\
&(iii)& \,
\bar{d}   \bar{d}   \to \bar{u}   \bar{u}   + \ell^+\ell^+.
\label{subproc}
\end{eqnarray}

The amplitudes are written using the following definitions of propagator
factors:
\begin{eqnarray}
1/A &=& \left[(p_a -p_c)^2-M_W^2\right] 
      \left[(p_b -p_d)^2-M_W^2\right]\cr
1/B &=& \left[(p_a -p_d)^2-M_W^2\right] 
      \left[(p_b -p_c)^2-M_W^2\right]\cr
1/\widetilde{A} &=& \left[(p_a +p_b)^2-M_W^2 +iM_W\Gamma_W\right] 
      \left[(p_c +p_d)^2-M_W^2+iM_W\Gamma_W\right]
\end{eqnarray}
\begin{eqnarray}
C &=&         {(p_a -p_c-p_e)^2-M_N^2}\cr
D &=&         {(p_a -p_c-p_f)^2-M_N^2}\cr
E &=&         {(p_a -p_d-p_e)^2-M_N^2}\cr
F &=&         {(p_a -p_d-p_f)^2-M_N^2}\cr
\widetilde{C} &=&         {(p_c +p_d+p_e)^2-M_N^2+iM_N\Gamma_N}\cr
\widetilde{D} &=&         {(p_c +p_d+p_f)^2-M_N^2+iM_N\Gamma_N}
\end{eqnarray}
The width of the heavy composite neutrino, $\Gamma_N$, is of course
a quantity which depends on the free parameters of the particular 
model that is being considered here, $|f|, \Lambda_{\hbox{c}}$ and  $M_N$, and
has been the object of discussion in the literature~\cite{djouadi,baur}.
Typically the width of  excited leptons (quarks) receives 
contributions from the  gauge interactions of Eq.~(\ref{leff}) 
and from contact terms arising 
from novel {\it strong} preon 
interactions~\cite{baur}~\footnote{It is to be 
noted however that these contact terms while contributing 
to the total
width of the excited neutrino cannot contribute to the production 
of LSD via the diagrams discussed in this work.}. 
In order to keep the numerical computations
of cross-sections presented in the following reasonably simple, 
a constant value of $\Gamma_N = 70$ GeV has been adopted, which is a somewhat
average value in the mass range considered.

Define also the quantities:
\begin{eqnarray}
s(m,n) &=& s(p_m,p_n) = \bar{u}_+(p_m) u_-(p_n)\, ,\nonumber \\
t(m,n) &=& t(p_m,p_n) = \bar{u}_-(p_m) u_+(p_n)\, ,
\end{eqnarray}
which are given by:
\begin{eqnarray}
s(m,n) & = & - 2\sqrt{E_mE_n}\ \  G_{mn}, \nonumber \\
t(m,n) & = & + 2\sqrt{E_mE_n}\ \  F_{mn},
\label{su}
\end{eqnarray}
with
\begin{eqnarray}
G_{mn} & = & \cos(\theta_m/2)\sin(\theta_n/2) 
\, e^{ +i(\phi_m-\phi_n)/2 } 
-\sin(\theta_m/2)\cos(\theta_n/2)
\, e^{ -i(\phi_m-\phi_n)/2} , \nonumber \\
F_{mn} & = &  (G_{mn})^*.
\label{gf}
\end{eqnarray}
Let the tensor $T_{\mu\nu}$  describe the virtual sub-process 
$W^\star W^\star \to \ell^+\ell^+$ (Fig.~\ref{fig2}a), 
while the tensor $\widetilde{T}_{\mu\nu}$ 
describes the virtual sub-process  
$(W^{\star})^+ \to \ell^+\ell^+ (W^{\star})^- $ appearing
in the diagram of Fig.~\ref{fig2}b.
$J_{a,c}$ and $\bar{J}_{b,d}$, are the quark (antiquark)
currents that couple in the t-channel to the 
virtual gauge bosons of the standard model (Fig.~\ref{fig2}a) while 
$\widetilde{J}_{a,b}$ and $\widetilde{J}_{c,d}^* $,
are the incoming and outgoing currents of the $q\bar{q}'$ pair
that couples in the s-channel to the W-bosons (Fig.~\ref{fig2}b).
\begin{eqnarray}
J^\mu_{a,c} 
& = & \bar{u}(p_c)\, \gamma^\mu\, \frac{1-\gamma_5}{2}\, u(p_a)\,\cr
\bar{J}^\mu_{b,d} 
& = & \bar{v}(p_b)\, \gamma^\mu\, \frac{1-\gamma_5}{2}\, v(p_d)\, \cr
\widetilde{J}^\mu_{a,b} 
& = & \bar{v}(p_b)\, \gamma^\mu\, \frac{1-\gamma_5}{2}\, u(p_a)\,\cr
({\widetilde{J}^{\mu}_{c,d}})^*
& = & \bar{u}(p_c)\, \gamma^\mu\, \frac{1-\gamma_5}{2}\, v(p_d)\ .
\end{eqnarray}
the amplitudes are (unitary gauge):

($i$) 
\underline{{$U_i U_j \to D_k D_l + \ell^+\ell^+$}}
\begin{equation}
{\cal M} = {\cal K} \,  \bar{u}(p_e)\,
\left\{ \, V_{U_iD_k} V_{U_jD_l} \, A \,  \left[ J^\mu_{(a,c)} \,   T_{\mu\nu}\,  
\, {J}^\nu_{(b,d)}\, \right]   
- V_{U_iD_l} V_{U_jD_k} \, B \, 
\Big[\, (p_c \leftrightarrow p_d)\, \Big] \right\}  \,  v(p_f) \, ;
\end{equation}

$(ii)$
\underline{\  $U_i \bar{D}_j   \to {D}_k   \bar{U}_l
+ \ell^+\ell^+ \,$}
\begin{eqnarray}
{\cal M}{(WW-\hbox{fusion})}& = &{\cal K} \, V_{U_iD_k} (V_{U_lD_j})^* \, 
\bar{u}(p_e)\, \left[ \, A\, J^\mu_{(a,c)} \,  T_{\mu\nu}\,  \bar{J}^\nu_{(b,d)}\, 
\right] \,  v(p_f)\, ;\cr
{\cal M}{(q\bar{q}'-\hbox{annihilation})}
& = &{\cal K} \, V_{U_iD_j} (V_{U_lD_k})^* \, 
\bar{u}(p_e)\, \left[ \, \widetilde{A}\, \widetilde{J}^\mu_{(a,b)} \,  
\widetilde{T}_{\mu\nu}\,  (\widetilde{J}^\nu_{(c,d)})^*\, 
\right] \,  v(p_f)\, ;
\end{eqnarray}

($iii$) 
\underline{{$\bar{D}_i \bar{D}_j \to \bar{U}_k \bar{U}_l 
+ \ell^+\ell^+$}}
\begin{equation}
{\cal M} = {\cal K} \,  \bar{u}(p_e)\,
\left\{ \, V_{U_kD_i}^* V_{U_lD_j}^* \, A \,  
\left[ \bar{J}^\mu_{(a,c)} \,   T_{\mu\nu}\,  
\, \bar{J}^\nu_{(b,d)}\, \right] - 
V_{U_lD_i}^* V_{U_kD_j}^* \, B 
\Big[ \, ( p_c \leftrightarrow p_d )\, \Big] \right\} \, v(p_f)\, ;
\end{equation}
where $U_i$ denotes a positively charged quark (up-type) while $D_i$ dentotes
a negatively charged one (down-type).
The quantities $V_{U_iD_j}$ are the elements
of the CKM mixing matrix. Of course the annihilation diagram of Fig. 2a comes in
only in quark-antiquark scattering. 
In prcesses ($i$) and ($iii$) the part of the amplitude depending on the 
factor $B$ is due to the diagrams obtained exchanging the
final state quarks.
In the framework of the effective Lagrangian 
c.f. Eq.~(\ref{leff}), as discussed in section II, 
it is found:
\begin{eqnarray}
T_{\mu\nu} & = &  \left[ \frac{\sigma_{\mu\rho}\sigma_{\nu\sigma}}
{C}
+ \frac{\sigma_{\nu\sigma}\sigma_{\mu\rho}}{D} 
\right] \frac{1-\gamma_5}{2} \, (p_a-p_c)^\rho (p_c-p_d)^\sigma \, ,
\cr
\widetilde{T}_{\mu\nu} & = &  \left[ \frac{\sigma_{\mu\rho}\sigma_{\nu\sigma}}
{\widetilde{C}}
+ \frac{\sigma_{\nu\sigma}\sigma_{\mu\rho}}{\widetilde{D}} 
\right] \frac{1-\gamma_5}{2} \, (p_a+p_b)^\rho (p_c+p_d)^\sigma\, ,
\cr
{\cal  K}  &= & \frac{g^4}{4} \left( \frac{f}{\Lambda_{\hbox{c}}}\right)^2
\, M_N\, .
\end{eqnarray}
Due to the chiral nature of the couplings involved, 
the calculation is particularly simple  if performed in 
the helicity basis~\cite{kleiss}. In the massless approximation 
only one helicity amplitude is non zero.
The following result is found:

\noindent ($i$) \underline{{$U_i U_j \to D_k D_l + \ell^+\ell^+$}}
\begin{eqnarray}
{\cal M}= 4\,  {\cal K} \, s(a,b) & \Bigg\{ & 
\phantom{+} V_{U_iD_k} V_{U_jD_l} \, A \,\, t(a,c) t(d,b) 
\left[\frac{ s(e,a) s(b,f)}{C}-\frac{s(f,a)s(b,e)}{D}
\right] \nonumber \\
& \, & - V_{U_iD_l} V_{U_jD_k} \, B \,\, t(a,d) t(c,b)
\left[\frac{s(e,a) s(b,f)}{E}-\frac{s(f,a)s(b,e)}{F}
\right]\Bigg\},
\label{ampsi}
\end{eqnarray}

\noindent ($ii$) \underline{{$U_i \bar{D}_j \to D_k \bar{U}_l + \ell^+\ell^+$}}
\begin{eqnarray}
\bullet & & (WW -\hbox{fusion}) : \nonumber \\ 
{\cal M}&=& + 4\, {\cal K} \, V_{U_iD_k} (V_{U_lD_j})^* 
\, A\, s(a,d)  t(a,c) t(d,b)  \Bigg\{ 
\left[\frac{ s(e,a) s(d,f)}{C}-\frac{s(f,a)s(d,e)}{D}
\right]\Bigg\},
\label{ampsii}
\end{eqnarray}
\begin{eqnarray}
\bullet & &(q\bar{q}'-\hbox{annihilation}) : \nonumber \\
{\cal M}&=& - 4\, {\cal K} \, V_{U_iD_j} (V_{U_lD_k})^* 
\, \tilde{A}\, t(a,b)  s(a,d) t(d,c)  \Bigg\{ 
\left[\frac{ s(e,a) s(d,f)}{\tilde{C}}-\frac{s(f,a)s(d,e)}{\tilde{D}}
\right]\Bigg\},
\label{ampsiii}
\end{eqnarray}

\noindent ($iii$) \underline{{$\bar{D}_i \bar{D}_j \to \bar{U}_k \bar{U}_l 
+ \ell^+\ell^+$}}
\begin{eqnarray}
{\cal M}= 4\,  {\cal K} \, s(c,d) & \Bigg\{ & 
\phantom{+} (V_{U_kD_i})^* (V_{U_lD_j})^*\, A \,\, t(a,c) t(d,b) 
\left[\frac{ s(e,c) s(d,f)}{C}-\frac{s(f,c)s(d,e)}{D}
\right] \nonumber \\
& \, & +(V_{U_lD_i})^* (V_{U_kD_j})^*\, B \,\, t(a,d) t(c,b)
\left[\frac{s(e,d) s(c,f)}{E}-\frac{s(f,d)s(c,e)}{F}
\right]\Bigg\}.
\end{eqnarray}
The above simple analytic form of the amplitudes is also very easy 
to implement in a code for numerical applications, since the quantities
$s(p_i,p_j)$ and  $u(p_i,p_j)$ are just functions of the energies and angles
of the particle's momenta, c.f. Equations~(\ref{su}) and (\ref{gf}). 

\section{Discussion and Results.}
Before giving details of numerical calculations of the signal 
cross-section and discussing the results one  should remind that 
there are processes of the standard model that also lead to LSD  
production and are thus sources of background to the signal. This question 
was already considered in refs~\cite{datta,dicus}. An immediate source of 
background comes from the subprocesses $uu \rightarrow dd W^{+} W^{+}$, 
 $u{\bar d} \rightarrow d{\bar u} W^{+} W^{+}$, ${\bar d}{\bar d}  
\rightarrow {\bar u} {\bar u} W^{+} W^{+}$ and similar ones involving 
higher-generation quarks and antiquarks, each $W$ subsequently decaying 
into $\ell \nu_{\ell}$. The corresponding overall reaction $ pp \rightarrow 
2\, \hbox{jets} \, \ell \nu_{\ell}\, \ell \nu_{\ell}$ 
can mimic the signal when the 
total missing $P_T$ carried away by the neutrinos is small. As shown in 
\cite{dicus}, that background can be most efficiently reduced to a percent of 
fb in LHC conditions, which will be shown to be at the same level of 
the signal, in some regions of the parameter space,
or even well below the latter in other regions. 
This background reduction is accomplished by limiting the missing 
$P_T$ of  neutrinos, that is, requiring a ``$P_T$-conservation'' 
which is actually a characteristic of the signal.

As also observed in~\cite{datta,dicus}, a copious and more 
dangerous source of 
standard-model background seems to be due to $t{\bar t}$ 
production from gluon and quark initial states. 
In that process, one has the decay chains $t\rightarrow b W^{+}$, 
$W^{+}\rightarrow
 \ell \nu_{\ell}$ on one side, and  ${\bar t}\rightarrow {\bar b } W^{-}$, 
${\bar b}\rightarrow {\bar c }\ell \nu_{\ell}$, $W^{-}\rightarrow q q'$ on 
the other side. For LHC conditions, that reaction leads to a total production 
of about $4\times 10^6$ LSD per year. 
Here again, a limitation of missing $P_T$ together with the condition
of large $P_T$ leptons
allow one to reduce substantially that background. The additional 
requirement of lepton isolation further reduces the background. 
But while the two requirements of missing $P_T$ limitation and lepton 
isolation will certainly eliminate two other similar backgrounds coming from 
direct $c{\bar c}$ and $b{\bar b}$ production, that of $t{\bar t}$ 
production seems to remain, according to~\cite{datta,dicus}, 
at a level which might 
jeopardize measurement of the signal at LHC. 

At this point, it is worth noticing that within the standard 
model one can observe in $p p$ collisions not only events with 
like-sign di-leptons of a given species ($e^{\pm}e^{\pm}$, 
${\mu}^{\pm}{\mu}^{\pm}$, ${\tau}^{\pm}{\tau}^{\pm}$), but 
also events with ``hybrid'' like-sign di-leptons (HLSD) such as 
$e^{\pm} {\mu}^{\pm}$, $e^{\pm}{\tau}^{\pm}$, ${\mu}^{\pm}{\tau}^{\pm}$, 
with practically the same production rate for all these events 
since the $W's$ decay into any $ \ell \nu_{\ell}$ final state at 
the same rate. Thus, one can get an idea on the amount of standard-model 
LSD background and eventually make appropriate subtraction by comparing, 
under given kinematical constraints, LSD production with HLSD production. 
At LHC, it would be most probably a comparison between ${\mu}^{\pm}
{\mu}^{\pm}$, ${\tau}^{\pm}{\tau}^{\pm}$ production and ${\mu}^{\pm}
{\tau}^{\pm}$ production. Said differently, once appropriate 
kinematical cuts performed, any significant difference between LSD 
production and HLSD production would signal lepton number violating 
processes like those 
here considered. However, let us remark that a no-deviation result could 
not rule out new physics models allowing for lepton mixing.

In any case let us remark that  an analysis of the background  
dedicated specifically to the LHC experimental conditions, and perhaps
more complete than that presented 
in~\cite{datta,dicus},
is necessary (including in particular a detailed calculation of the 
amplitude of the processes involved), and will be 
the matter of a forthcoming work.
Here the estimate of the background given in~\cite{dicus} is assumed:
\begin{equation}
\sigma_{background} = 3 \times 10^{-2} \, \hbox{fb}.
\label{bg}
\end{equation}
In order to compare the signal cross-section with Eq.(\ref{bg}) 
 kinematical cuts as discussed in~\cite{dicus} are used.
The following selection criteria are needed 
in order to ensure lepton and jet identification:
\begin{eqnarray}
|\eta_{lep} | &< & 4 \qquad p_T(lep) > \phantom{0}5\, \,  \hbox{GeV},\cr
|\eta_{jet} | &<& 4 \qquad p_T(jet) > 20\, \, \hbox{GeV}. 
\end{eqnarray}

The signal cross-sections are obtained by folding the square of the amplitudes
with the four-particle phase-space and the parton distribution functions:
\begin{eqnarray}
d\sigma = \int dx_a dx_b\, \frac{1}{1+\delta_{ij}}&\,& 
\left[ f_i(x_a,Q^2)   f_j(x_b,Q^2) + x_a \leftrightarrow x_b \right] \times \cr
&\, &\frac{1}{2\hat{s}}\, 
|{\cal M } |^2\, (2\pi)^4 \, \delta^4 (p_a+p_b -\sum_{m=1}^4 p_m)\,
\, \frac{1}{2}\, 
(1-\frac{\delta_{kl}}{2})\, \prod_{n=1}^4 \, \frac{d^3 \bm{p}_n}{(2\pi)^3 2E_n},
\label{cross}
\end{eqnarray}
where $\hat{s} = x_a x_b S$ is the squared center of mass energy of the parton 
collision and the factor $(1/2)(1-\delta_{kl}/2)$ accounts for the presence 
of the two identical fermions ($\ell^+\ell^+$) and the  possibly identical quarks 
$U_k,U_l $ ($\bar{U}_k,\bar{U}_l$) in the final state. 
The distribution functions are those of Set 1.1 of 
Duke-Owens (updated version of Set 1) 
as described in~\cite{owens} with $\Lambda_{QCD}=177 $ MeV/c. 
$\sqrt{S} = 14 $ TeV has been used 
while the scale $Q^2$ is  fixed at the value 
$Q^2=\hat{s}$.  
With a proper choice of the transverse axis the phase-space reduces 
to a nine-dimensional integration that is performed with the well 
known VEGAS~\cite{lepage} routine which is based on a Monte-Carlo algorithm. 
This allows  easy   implementation of  
kinematical cuts as described above. 
As regards the $u\bar{d}$ process the interference between the
WW-fusion and annihilation mechanisms is naturally taken into account
since the two (complex) amplitudes are summed before squaring.

In Figs.(\ref{lhc_a} \& \ref{lhc_b}) the integrated cross-section 
with the parameter $|f|=1 $, i.e. $\sigma_1 = \sigma (|f|=1)$ is given.
As  ${\cal M} \propto f^2 $,  the total cross
section for other values of $|f|$ can be easily recovered 
($\sigma = |f|^4 \times \sigma_1 $).
Keeping fixed $|f|=1$ there are  other two parameters  on which our signal 
rate is dependent: $\Lambda_{\hbox{c}}$ and $M_N$. In order to sample
different regions of the parameter space two cases have been considered.
Case $(a)\, \Lambda_{\hbox{c}}=1$ TeV, and case $(b)\, \Lambda_{\hbox{c}}=M_N$.

Case $(a)$ is shown in Fig.(\ref{lhc_a})  
where cross-sections corresponding 
to  the three subprocesses of the first quark 
family, c.f. Eq.(\ref{subproc}), 
are plotted versus the mass of the excited Majorana neutrino 
$M_N$~(Fig.~\ref{lhc_a}a,\ref{lhc_a}b,\ref{lhc_a}c).
Since the subprocess 
$\bar{d} \bar{d} \to \bar{u} \bar{u} + \ell^+\ell^+ $
is weighted by sea-quark distribution functions it is
is totally negligible relative to the other two.
In  Fig.(\ref{lhc_a}d) the {\it total cross section} is plotted versus
$M_N$  including  contributions
from other subprocesses with  second generation quarks, ($c,s$) as described 
in the appendix. Some of these subprocesses 
are however weighted by off-diagonal elements of the CKM mixing matrix
and therefore give only small corrections.
The shape of the curves as a  function of $M_N$ is clearly understood since
the only dependence on the new parameters is of the 
type~\footnote{It should be remarked that this is only true within 
the approximation of a constant width $\Gamma_N$ for the heavy neutrino. 
Taking into account the dependence of $\Gamma_N$ with the new physics 
parameters $|f|, M_N$ and $\Lambda_{\hbox{c}}$ (and those pertaining to 
contact terms)  could modify, to some extent, the contibution of the 
quark-antiquark scattering. However, as pointed out 
in ref.~\cite{baur}, $\Gamma_N$ receives the largest 
contribution from contact terms which are independent of $|f|$.
}:
\begin{equation}
\sigma \sim \left(\frac{|f|}{\Lambda_{\hbox{c}}}\right)^4 \, M_N^2 \, 
\int \sum_K \, \frac{1}{(K^2 - M_N^2)^2 + \theta (K^2) (M_N \Gamma_N)^2}
\label{behav}
\end{equation}
where $K$ are different momenta flowing in the Majorana propagator. Thus
in case $(a)$, $\sigma \to 0 $ as $M_N\to 0$ while $\sigma \sim M_N^{-2} $
as $M_N \to \infty $, and there is an intermediate region with a maximum.
There is a mass interval from $M_N = 250 $ GeV up to $M_N \approx 3 $ TeV
where $\sigma_1$ is bigger than the lowest measurable 
cross-section of $10^{-2}$ fb
that corresponds to one event per year given the luminosity 
${\cal L}_0 = 100 $ fb$^{-1}$ (integrated over one year) planned at LHC.
For example the total signal cross-section $\sigma_1$ is at most about 
$ 5 \times 10^{-2}$ fb,
which is at the same level (though bigger) of the background (Eq.~\ref{bg}),
and would only give five  events per year. 
It seems therefore that,  {\it with this particular choice
of parameters} 
[case $(a)\, \Lambda_{\hbox{c}}=1$ TeV, $|f| = 1$], 
the lepton number violating 
signal due to the  composite Majorana neutrino 
would hardly be measurable, 
unless a better set of kinematical cuts is found that enhances 
the absolute value of the signal rate while reducing still
further the background. However one should keep in mind the
dependence on the parameter $|f|$, which in Fig.(\ref{lhc_a}) has
been fixed to  $|f|=1 $. Since the signal cross section is proportional
to  $ |f|^4 $ even a slightly larger value of $|f|$ could increase
sensibly the signal cross-section. 

Case $(b)$ is shown in Fig.(\ref{lhc_b}) with the same notation as 
in Fig.(\ref{lhc_a}). Again the subprocess 
$\bar{d} \bar{d} \to \bar{u} \bar{u} + \ell^+\ell^+ $ [Fig.(\ref{lhc_b}c)]
is totally negligible relative to the other two.
The different shape of the cross section $\sigma_1$ as a function
of $M_N$ is of course due to the choice $\Lambda_{\hbox{c}} = M_N$,
which according to Eq.(\ref{behav}) gives roughly $\sigma_1 \sim M_N^{-6}$
as $M_N \to \infty$ while $\sigma_1 \sim M_N^{-2} $ as $M_N \to 0$.
Thus $\sigma_1$ is strongly enhanced respect to case $(a)$ for values
of $M_N < 1 $ TeV, while for $M_N > 1 $ TeV it will be severely decreased.
The cross-section $\sigma_1$ will be measurable in the mass interval
$M_N=250$ GeV (400 events/year) up to $M_N \approx 1.4 $ TeV (1 event/year).
This portion of the parameter space has therefore the potential of giving 
rise to a signal with a substantially higher number of events respect to
the background, at least up to $M_N=850 $ GeV (10 events/year).

Finally it should be remarked that the discussion so far has been
quite general with respect to the lepton flavour and applicable 
to all three of them but, 
(LSD $=\ell^{\pm}\ell^\pm,\, \ell= e,\mu,\tau$)   
at the LHC, muons will be the leptons most easily detected 
while the other lepton flavours 
will be detectable but with lower efficiencies~\cite{servoli}.  
For this reason the numerical results
presented here refer to only one lepton generation.
 
\section{Comparing the LHC vs the GENIUS potential}

This section contains a comparative discussion of the constraints
on the parameters $|f|, M_N ,\Lambda_{\hbox{c}}$ that could be derived
by the non observation of the $L$-violating signals discussed in the 
previous section at the {\it high-energy} 
LHC experiments as opposed to those deriving from the non-observation
of {\it low-energy} neutrinoless double beta decay experiments, 
present (Heidelberg-Moscow) and next-generation (GENIUS).
The new ${\beta\beta}_{0\nu}$ GENIUS experiment, 
(GErmanium-detectors  in liquid
NItrogen as shielding in an Underground  Setup)~\cite{hellmig}, 
has the potential to improve by orders of magnitude the lower bound
on the $\beta\beta_{0\nu}$ decay half-life. Monte-Carlo simulations
have shown that in one (four) year(s) of measurement the lower bound
will  be increased respectively to~\cite{hellmig,private}:
\begin{eqnarray}
T_{1/2}^{0\nu} &>& 5.8 \times 10^{27}\, \, yr, \qquad \hbox{[one year]}
\nonumber \cr 
T_{1/2}^{0\nu} &>& 2.3 \times 10^{28}\, \, yr. \qquad \hbox{[four years]} 
\end{eqnarray}
Figures~\ref{genius_a} \& \ref{genius_b} show the upper bound on the
parameter $|f|$ as function of the heavy neutrino mass $M_N$
for the two cases $(a)$   and $(b)$ defined in 
the previous section.  
The curves concerning the $\beta\beta_{0\nu}$ bound 
are based on formulas that can be found in~\cite{opring}
which relative to Eq.~(\ref{constraint_num}) above include small 
correction terms of order ${\cal O}(M_W/M_N)$.
The LHC curves are found using the numerical cross-sections 
presented in the previous section, requiring $10$ events/year as a criterion
for discovery of the $L$-violating signal, and 
assuming an integrated luminosity
of ${\cal L}_0 = 100$ fb$^{-1}$ as before. Thus non-observation of the signal
at LHC means that
$|f|^4\, \sigma_1(M_N,\Lambda_{\hbox{c}})\, {\cal L}_0 < 10 $,
which is translated into  a constraint on
 $|f|$ that is the corresponding LHC upper bound
to that in Eq.~(\ref{constraint_num}) from $\beta\beta_{0\nu}$:
\begin{equation}
|f| < \left(\frac{10}{{\sigma_1}{\cal L}_0}\right)^{1/4}\, .
\end{equation}

From Figures~\ref{genius_a} \& \ref{genius_b} 
one can infer {\it lower bounds} on the composite
neutrino mass (or equi\-va\-len\-tely the compositeness scale)
by {\it assuming } the dimensionless coupling $| f | \sim {\cal O}(1) $.
For case $(a)$, Fig.~\ref{genius_a}, one obtains  
the bounds shown in Table I while in Table II  the corresponding 
bounds for case $(b)$, Fig.~\ref{genius_b}, are given.
One comment is in order here. The LHC curve in Fig.~\ref{genius_a} has
a different behaviour for $M_N < 1 $  TeV as 
compared to those of the $\beta\beta_{0\nu}$. This is due to the fact that
as $M_N \to 0$, $\sigma_1 \to 0$ and thus the LHC upper bound on $|f|$
becomes weaker and weaker. This does not happen in the $\beta\beta_{0\nu}$
whose squared amplitude behaves as $|{\cal M}_{\beta\beta_{0\nu}}|^2 \sim
M_N^{-2}$~\cite{op3} and at lower 
masses gives a bigger effect and therefore a stronger constraint.
It is for this reason that Table I, for case $(a)$, 
does not show a lower bound on $M_N$ for LHC.
In case $(b)$  
if $|f| \sim {\cal O }(1)$ GENIUS--($1$ yr) can exclude 
Majorana composite neutrinos up to a mass of $M_N \sim 700 $ GeV, 
while LHC and GENIUS--($4$ yr) can go up to about  
$850$ GeV. 
It is important to realize that the {\it non-accelerator, low-energy}, 
GENIUS-4yr experiment has the potential 
to probe the compositeness scale into the TeV region. 

At this point the reader should be made aware that
investigations of the same type of effective Lagrangians for compositeness
within the context of LHC experiments have  already been reported 
in the literature. In particular while   
the production of  excited quarks at LHC has been investigated both 
via magnetic type  gauge (G) interactions 
and contact terms (CT)~\cite{djouadi},
the production of excited {\it leptons} 
has however been considered {\it only through} CT and a 
mass sensitivity of  up to about $ 4 - 5 $ TeV 
is found~\cite{djouadi}.  
This work is therefore the first report 
concerning excited leptons at LHC within
the context of magnetic type gauge interactions, and,
while the discovery limit
derived for contact terms~\cite{op2,djouadi} 
cannot be directly compared with 
the constraints derived  in~\cite{op1,op2,op3} 
from the non-observation of $\beta\beta_{0\nu}$ (that were based on gauge
interactions G), the discovery limit for LHC reported here
($M_N$ up to $850$ GeV)
can be directly compared with that of $\beta\beta_{0\nu}$ as done explicitly
in Table II and Figures~\ref{genius_a} and  \ref{genius_b}.

Finally it is worthwhile to note
the complementary role that accelerator (LHC) and non-accelerator
experiments (GENIUS) can have. 
Figures~\ref{genius_a} and \ref{genius_b} show explicitly 
that, in both cases $(a)$ and $(b)$, 
while for low masses the $\beta\beta_{0\nu}$
bound  is more restrictive there is always a crossing point
where the LHC constraint becomes stronger, though of the same order
of magnitude.

\section{Conclusions}
In this work the production of Like Sign Di-leptons (LSD) via the 
exchange  of a heavy composite Majorana neutrino in $pp$ collisions
has been studied in detail at LHC energies. The coupling of the Majorana
neutrino is assumed to be a gauge interaction of the magnetic moment type 
($\sigma_{\mu\nu}$). The helicity amplitudes have been presented and 
the resulting cross-sections within kinematical cuts, needed to
suppress the SM background down to the fb level, 
are reported. Regions of the parameter space
are pinned down where the signal is well above the estimated 
background ($\Lambda_{\hbox{c}} = M_N , |f| \sim 1 , M_N < 850 $ GeV).
However a study of the background
specifically dedicated to the LHC experimental conditions would
certainly be of help towards a better understanding of the lepton
number violating processes discussed here.
The comparison of the LHC potential with respect to observing 
$L$-violating processes with that of the new generation 
of the {\it non-accelerator type}  
$\beta\beta_{0\nu}$ experiment,  GENIUS, shows how the two
approaches, {\it high- vs. low-energy}, do play a complementary role.

The approach developed here to discuss LSD production via 
composite Majorana neutrinos at LHC is  being
extended to other models of physics beyond the SM 
which provide $L$-violating interactions. 
The results of these analysis will be reported elsewhere.
 
One final remark is to be added concerning the interplay of {\it low-} vs. 
{\it high-energy} facilities with respect to the study of 
lepton number violation. 
The class of diagrams 
that give rise to $\Delta L = \pm 2$ processes discussed in this work 
could also trigger lepton number violating 
rare Kaon decays 
such as $ K^+ \to \pi^- e^+ e^+ $. At the Frascati
$\Phi$-factory, DA${\Phi}$NE~\cite{daphne} (presently under commissioning), 
these decays could either be observed or, 
otherwise,  the corresponding bounds on the 
branching ratios  are susceptible to be strengthened. 
The current bound
on the branching ratio for the  ($\Delta L = - 2$) $K^+$ decay is 
$Br( K^+ \to \pi^- e^+ e^+) < 1.0\times 10^{-8}$~\cite{pdg}, 
while the sensitivity of the KLOE experiment~\cite{kloe} to be performed 
at DA${\Phi}$NE could reach the level of $10^{-9}$; the KLOE experiment
might thus provide  insights on  lepton number violating interactions 
beyond the standard model. 
Work along these lines is in progress.
 
\acknowledgments
The authors would  like to thank M.~Espirito-Santo, of 
the DELPHI Collaboration, for providing the data histogram plotted 
in Figure 3.

O.~P. would like to acknowledge useful discussions with 
the experimental colleagues of the Compact Muon Solenoid (CMS) Collaboration,
L.~Servoli, P.~Cenci G.~M.~Bilei and A.~Nappi.

C.~C. acknowledges support by a  grant from the Istituto Nazionale 
di Fisica Nucleare of Italy (INFN), that allowed his stay in Perugia, 
where this work was completed.  He is also indebted to Z. Ajaltouni 
(ALEPH Collaboration), F.~Fleuret and S. Jan (ATLAS Collaboration) 
for useful discussions.

This project is partially supported by the EEC-TMR Program, 
Contract N. CT98-0169.  

\appendix

\section{List of Parton subprocesses}
A list of  all subprocesses leading to the production of LSD within the 
first two families of quarks is:
\begin{itemize}

\item[{($i$)}] quark scattering, 
$U_i U_j \to D_k D_l + \ell^+\ell^+$, 
\subitem
\begin{tabular}{||c|c||}
\hline\hline
\ $(k=l)$\ & \ $(k\neq l)$\ \cr\hline
\ \bm{u u  \to d d \, \, [ ss\, ] + \ell^+\ell^+}\  & \ \bm{ u u \to d s \, 
+ \ell^+\ell^+}\ \cr
\hline
\ $c c  \to s s \, \, [ dd\, ] + \ell^+\ell^+$\  & \ $ c c \to d s \,
+ \ell^+\ell^+$\ \cr
\hline
\ $u c  \to s s \, \, [ dd\, ] + \ell^+\ell^+$\  & \ \bm{ u c \to d s \, 
+ \ell^+\ell^+}\ \cr
\hline\hline
\end{tabular}
\item[{($ii$)}] quark antiquark scattering ($U_i \bar{D}_j
                \to D_k \bar{U}_l + \ell^+\ell^+$):
\subitem
\begin{tabular}{||c||}
\hline\hline
\ \bm{u \bar{d} \to d \bar{u} \,\, [d \bar{c},s \bar{u},s\bar{c}\, ] + \ell^+\ell^+}\ \cr 
\hline
\ \bm{u \bar{s} \to d \bar{c} \,\, [d \bar{u},s \bar{u},s\bar{c}\, ] + \ell^+\ell^+}\ \cr 
\hline
\ $c \bar{s} \to s \bar{c} \,\, [s \bar{u},d \bar{c},d\bar{u}\, ] + \ell^+\ell^+$\ \cr
\hline 
\ $c \bar{d} \to s \bar{u} \,\, [s \bar{c},d \bar{c},d\bar{u}\, ] + \ell^+\ell^+$\ \cr
\hline\hline
\end{tabular}
\item[($iii$)]  anti-quark scattering  
($\bar{D}_i \bar{D}_i \to \bar{U}_k \bar{U}_l + \ell^+\ell^+$) :
\subitem 
\begin{tabular}{||c|c||}
\hline\hline
\ $ (k=l)$ & \ $(k \neq l)$ \cr
\hline
\ $\bar{d} \bar{d}  \to \bar{u} \bar{u} \, \, [ \bar{c} \bar{c} \, ] + \ell^+\ell^+$\  &
\ $\bar{d} \bar{d}  \to \bar{u} \bar{c} \, + \ell^+\ell^+$\  \cr 
\hline
\ $\bar{s} \bar{s}  \to \bar{c} \bar{c} \, \, [ \bar{u} \bar{u} \, ] + \ell^+\ell^+$\  &
\ $\bar{s} \bar{s}  \to \bar{u} \bar{c} \,   + \ell^+\ell^+$\ \cr
\hline 
\ $\bar{d} \bar{s}  \to \bar{u} \bar{u} \, \, [ \bar{c} \bar{c} \, ] + \ell^+\ell^+$\  &
\ $\bar{d} \bar{s}  \to \bar{c} \bar{u} \,   + \ell^+\ell^+$\ \cr
\hline\hline
\end{tabular}
\end{itemize}

Numerical results reported in Figures~\ref{lhc_a}d and \ref{lhc_b}d 
contain contributions from some of the processes listed above.
Processes initiated by two {\it sea} partons and not receiving contribution
from the annihilation diagram have not been considered
since  Fig.~\ref{lhc_a}c and \ref{lhc_b}c show that they are clearly
negligible.
As regards quark scattering only two cases have been considered;
sub-processes initiated by $uu$  and  
$u c$ collisions, {\it i.e.} with at least one $u$-quark in the initial state:
\begin{itemize}
\item \underline{$uu$ initiated sub-processes}
\subitem {\bf -} the processes
$uu \to dd +\ell^+ \ell^+ $ ,  $uu \to ds +\ell^+ \ell^+ $ and 
$uu \to ss +\ell^+ \ell^+ $  are factorized as follows:
\begin{equation}
|{\cal M}_{uu-initiated}|^2  =  
\left [ 1  
+2\left | \frac{V_{us}}{V_{ud}}\right |^2 
+ \left | \frac{V_{us}}{V_{ud}}\right |^4 \right ] \, \times \,
|{\cal M}_{uu \to dd +\ell^+ \ell^+} |^2 \, , 
\end{equation}
the additional factor of $2$, in the equation above, accounts for the fact
that the process  $uu \to ds +\ell^+ \ell^+ $ 
does not contain identical quarks in the final state as opposed to the processes
$uu \to dd +\ell^+ \ell^+ $ and $uu \to ss +\ell^+ \ell^+ $ and thus for it 
Eq.~\ref{cross} applies with $ k\ne l $.
\end{itemize}
Quark-antiquark scattering sub-processes have been divided into:
\begin{itemize}
\item \underline{$ u \bar{d}$ collisions}
\subitem {\bf -}
the processes $u \bar{d} \to\, [d \bar{u},s \bar{u},d\bar{c}\, ] + \ell^+\ell^+$ 
are factorized as:
\begin{equation}
|{\cal M}_{u\bar{d}-initiated}|^2  =
\left [ 1 + \left | \frac{V_{us}}{V_{ud}}\right |^2 
+ \left | \frac{V_{dc}}{V_{ud}}\right |^2 \right ] \, \times \,
|{\cal M}_{u\bar{d} \to d\bar{u} +\ell^+ \ell^+} |^2\, ; 
\end{equation}
\subitem {\bf -}
the process $u\bar{d}\to s\bar{c} +\ell^+ \ell^+ $,
that will turn out to be numerically the most important, 
between those containing second  family partons, does not factorize as above
due to the fact that the $WW$-fusion and 
the annihilation diagram come in with different
factors of the elements of the CKM matrix:
\begin{equation}
|{\cal M}_{u\bar{d} \to s \bar{c}+\ell^+ \ell^+}|^2 = \left | 
\frac{V_{us}V_{dc}}{V_{ud}^2} 
{\cal M}_{u\bar{d} \to d\bar{u} +\ell^+ \ell^+}^{(WW-fusion)} + 
\frac{V_{cs}}{V_{ud}}
{\cal M}_{u\bar{d} \to d\bar{u} +\ell^+ \ell^+}^{(u\bar{d}-annihil.)}
\right |^2 \, ,
\end{equation}
(see Eq.~23).

\item \underline{$u\bar{s}$ collisions}
\subitem {\bf -}
the processes $u \bar{s} \to \, 
[s \bar{u},d \bar{u},s\bar{c}\, ] + \ell^+\ell^+$ can be factorized as :
\begin{equation}
|{\cal M}_{u\bar{s}-initiated}|^2  
= \left (\frac{V_{us}}{V_{ud}} \right )^4
\left [ 1 + \left | \frac{V_{us}}{V_{ud}}\right |^2 
+ \left | \frac{V_{cs}}{V_{us}}\right |^2 \right ]\, \times \,
|{\cal M}_{u\bar{d} \to d\bar{u} +\ell^+ \ell^+} |^2,
\end{equation}
and using the fact the within the set of parton densities used here
(set 1.1 of Owens~\cite{owens}), 
$\bar{u}(x)=\bar{d}(x)=\bar{s}(x)$, the cross section for $u\bar{s}$
initiated collisions can be simply obtained from $\sigma_1{(u\bar{d} \to d \bar{u}
+\ell^+ \ell^+)}$ by multiplying it with the above CKM factor wich is 
$0.054 \approx 5\% $;
\subitem {\bf -}
the process ${u\bar{s} \to d\bar{c} +\ell^+ \ell^+}$ does not factorize
as in the above equation and must be considered separately (it is shown in 
Fig.~\ref{lhc_a2}):
\begin{equation}
{\cal M}_{u\bar{s} \to d\bar{c} +\ell^+ \ell^+}=
\left | \frac{V_{cs}}{V_{ud}} 
{\cal M}_{u\bar{d} \to d\bar{u} +\ell^+ \ell^+}^{(WW-fusion)} + 
\frac{V_{us}V_{dc}}{(V_{ud})^2}
{\cal M}_{u\bar{d} \to d\bar{u} +\ell^+ \ell^+}^{(u\bar{d}-annihil.)}
\right |^2 \, ;
\end{equation}
\item \underline{$c\bar{s}$ collisions}
\subitem {\bf -}
the processes $c \bar{s} \to \, 
[s \bar{c},s \bar{u},d\bar{c}\, ] + \ell^+\ell^+$ can be factorized as :
\begin{eqnarray}
|{\cal M}_{c\bar{s}-initiated}|^2  
&=& \left (\frac{V_{cs}}{V_{ud}} \right )^4
\left [ 1 + \left | \frac{V_{us}}{V_{cs}}\right |^2 
+ \left | \frac{V_{cd}}{V_{cs}}\right |^2 \right ]\, \times \,
|{\cal M}_{u\bar{d} \to d\bar{u} +\ell^+ \ell^+} |^2, \nonumber \cr
&=& \left [ 1 + \left | \frac{V_{us}}{V_{cd}}\right |^2 
+ \left | \frac{V_{cs}}{V_{cd}}\right |^2 \right ]\, \times \,
|{\cal M}_{c\bar{s} \to d\bar{c} +\ell^+ \ell^+} |^2,
\end{eqnarray}
\subitem {\bf -}
finally the process $c\bar{s} \to d\bar{u} +\ell^+ \ell^+$ has to be considered
separately:
\begin{equation}
|{\cal M}_{c\bar{s} \to d\bar{u} +\ell^+ \ell^+} |^2 =
\left | \frac{V_{cd}V_{us}}{V_{ud}^2} 
{\cal M}_{u\bar{d} \to d\bar{u} +\ell^+ \ell^+}^{(WW-fusion)} + 
\frac{V_{cs}}{V_{ud}}
{\cal M}_{u\bar{d} \to d\bar{u} +\ell^+ \ell^+}^{(u\bar{d}-annihil.)}
\right |^2 \, ;
\end{equation}
\item \underline{$c\bar{d}$ collisions}
\subitem {\bf -}
the processes $c \bar{d} \to \, 
[s \bar{c},d \bar{u},d\bar{c}\, ] + \ell^+\ell^+$ can be factorized as :
\begin{eqnarray}
|{\cal M}_{c\bar{d}-initiated}|^2  
&=& \left (\frac{V_{cd}}{V_{ud}} \right )^2
\left [ 1 + \left | \frac{V_{cd}}{V_{ud}}\right |^2 
+ \left | \frac{V_{cs}}{V_{ud}}\right |^2 \right ]\, \times \,
|{\cal M}_{u\bar{d} \to d\bar{u} +\ell^+ \ell^+} |^2, \, \nonumber \cr
&=& \left [ 1 + \left | \frac{V_{cd}}{V_{cs}}\right |^2 
+ \left | \frac{V_{ud}}{V_{cs}}\right |^2 \right ]\, \times \,
|{\cal M}_{c\bar{s} \to d\bar{c} +\ell^+ \ell^+} |^2,
\end{eqnarray}
\subitem {\bf -}
finally the process $c\bar{d} \to s\bar{u} +\ell^+ \ell^+$ has to be considered
separately:
\begin{equation}
|{\cal M}_{c\bar{d} \to s\bar{u} +\ell^+ \ell^+} |^2 =
\left | \frac{V_{cs}}{V_{ud}} 
{\cal M}_{u\bar{d} \to d\bar{u} +\ell^+ \ell^+}^{(WW-fusion)} + 
\frac{V_{cd}V_{us}}{V_{ud}^2}
{\cal M}_{u\bar{d} \to d\bar{u} +\ell^+ \ell^+}^{(u\bar{d}-annihil.)}
\right |^2 \, ;
\end{equation}
\end{itemize}

Finally the amplitude of the process $uc \to ds +\ell^+ \ell^+ $
although
weighted by only one u-quark distribution function contains a graph 
multiplied by 
diagonal elements of the CKM
matrix  ($\propto V_{ud}^2V_{cs}^2 $) and turns out to yield a contribution
comparable to that of the  $q\bar{q}'$ sub-processes 
described above (see Fig.~\ref{lhc_a2}).

Eqs.~(A1-A5) have been adopted to estimate the contribution of 
the subprocesses due to second family partons. Note that 
in numerical computations, 
the complex phases of the elements of the CKM mixing matrix
have been neglected, assuming $V_{ij} = |V_{ij}|$, as only the first
two generations are being considered here.

\section{Square of Amplitudes}
For the convenience of the reader interested in numerical applications 
the square of the amplitudes of the WW fusion mechanism 
is given here expressed in terms of the 
particles' momenta scalar products.
In the numerical calculations it has been checked 
that one obtains an agreement 
of 1 part in $10^5$ between  this way of calculating the square of the
amplitudes and the other consisting in writing down complex amplitudes 
and numerically taking the square of the absolute value.

Defining the quantities $K_{i} (i=1,2,3)$ by
\begin{equation}
\sum_{\hbox{pol}}|{\cal M}_i|^2 = 512 \, {\cal F}_{CKM}^{(i)} 
\, {\cal K}^2 \, K_{i}
\end{equation}
they are explicitly:

\noindent ($i$) \underline{{$U_i U_j \to D_k D_l + \ell^+\ell^+$}}

\begin{eqnarray}
K_{i} = p_a.p_b\, 
\Bigg\{ &+&A^2 \, p_a.p_c\, p_b.p_d \, \Bigg[+ \frac{p_a.p_e \, p_b.p_f}{C^2} 
+
\frac{p_a.p_f\, p_b.p_e}{D^2}-\frac{L(p_a,p_e,p_b,p_f)}{C D}\Bigg]\cr
&+&B^2  \, p_a.p_d\, p_b.p_c \, \Bigg[ + \frac{p_a.p_e \, p_b.p_f}{E^2} 
+
\frac{p_a.p_f\, p_b.p_e}{F^2}- \frac{L(p_a,p_e,p_b,p_f)}{E F}\Bigg]\cr
&-&AB\, \Bigg\{ L(p_a,p_c,p_b,p_d)
\Bigg[\frac{p_a.p_e\, p_b.p_f}{CE}+\frac{p_a.p_f\, p_b.p_e}{DF}\cr
& &\phantom{xxxxxxxxxxxxxxxxxxxxxxxx} -\frac{1}{2}\, L(p_a,p_e,p_b,p_f) 
\left(\frac{1}{CF}+\frac{1}{DE}\right)\Bigg]\cr 
& &\phantom{xxxx} -\frac{1}{2}\, 
\epsilon(p_a,p_b,p_c,p_d).\epsilon(p_a,p_b,p_e,p_f)
\left(\frac{1}{CF}-\frac{1}{DE}\right)\Bigg\}\Bigg\}
\end{eqnarray}
\noindent ($ii$) \underline{{$U_i \bar{D}_j \to D_k \bar{U}_l + \ell^+\ell^+$}}
\begin{equation}
K_{ii} = p_a.p_d\, p_b.p_d\, p_c.p_a \, A^2
\Bigg\{ + \frac{p_e.p_a \, p_f.p_d}{C^2} 
+
\frac{p_f.p_a\, p_e.p_d}{D^2} - \frac{L(p_e,p_a,p_f,p_d)}{CD}\Bigg\}
\end{equation}

\noindent ($iii$) \underline{{$\bar{D}_i \bar{D}_j \to \bar{U}_k \bar{U}_l 
+ \ell^+\ell^+$}}

\begin{eqnarray}
K_{iii} = p_c.p_d\, 
\Bigg\{ &+&A^2 \, p_a.p_c\, p_b.p_d \, \Bigg[+ \frac{p_c.p_e \, p_f.p_d}{C^2} 
+
\frac{p_c.p_f\, p_e.p_d}{D^2}-\frac{L(p_c,p_e,p_d,p_f)}{C D}\Bigg]\cr
&+&B^2  \, p_a.p_d\, p_b.p_c \, \Bigg[ + \frac{p_c.p_f \, p_e.p_d}{E^2} 
+
\frac{p_c.p_e\, p_f.p_d}{F^2}- \frac{L(p_c,p_e,p_d,p_f)}{E F}\Bigg]\cr
&-&AB\, \Bigg\{ L(p_a,p_c,p_b,p_d)
\Bigg[\frac{p_c.p_e\, p_f.p_d}{CF}+\frac{p_c.p_f\, p_e.p_d}{ED}\cr
& &\phantom{xxxxxxxxxxxxxxxxxxxxxxxx} -\frac{1}{2}\, L(p_e,p_c,p_f,p_d) 
\left(\frac{1}{CE}+\frac{1}{DF}\right)\Bigg]\cr 
& &\phantom{xxxx} -\frac{1}{2}\, 
\epsilon(p_a,p_b,p_c,p_d).\epsilon(p_e,p_f,p_c,p_d)
\left(\frac{1}{CE}-\frac{1}{DF}\right)\Bigg\}\Bigg\}
\end{eqnarray}
with $L(p_a,p_b,p_c,p_d) = p_a.p_b\, p_c.p_d + p_a.p_d\, p_b.p_c 
- p_a.p_c\, p_b.p_d$\ .

\begin{table}
\caption{Lower bound on $M_N$ for case $(a)$ 
[$\Lambda_{\hbox{c}} = 1 \hbox{ TeV}$ , $| f | = 1 $]. 
The bounds are derived from the non observation
of neutrinoless double beta decay ($\beta\beta_{0\nu}$) at the current
(Heidelberg-Moscow) experiment and for the prospected GENIUS experiment 
after 1 and 4 years of running~\protect\cite{hellmig}.
At LHC non observation of the LSD signal would not imply a lower
bound on the composite neutrino mass because of the different shape
of the exclusion plot. See Fig.~\protect\ref{genius_a}.
}
\begin{tabular}{lll}
Experiment & Exp. constraint  
& Lower Bound on $M_N \, $ 
(GeV)\cr
\hline   
Heidelberg-Moscow & $ T_{1/2} > 7.4 \times 10^{24} $ yr & $
M_N > \sim \phantom{0}10$ \cr
GENIUS $1$ yr & $T_{1/2} > 6.0 \times 10^{27} $ yr & $M_N > \sim 350$ \cr
GENIUS $4$ yr & $T_{1/2} > 2.3 \times 10^{28} $ yr & $M_N > \sim 700$ \cr
LHC & $N_{events}  < 10 $ &  $ ------ $\cr
\end{tabular}
\end{table}

\begin{table}
\caption{Lower bound on $M_N$ for case $(b)$ 
[$\Lambda_{\hbox{c}} = M_N$ , $| f | = 1 $]. 
The bounds are derived from the non observation
of neutrinoless double beta decay ($\beta\beta_{0\nu}$) at the current
(Heidelberg-Moscow) experiment and for the prospected GENIUS experiment 
after 1 and 4 years of running~\protect\cite{hellmig} and from
non observation of the LSD signal at LHC 
(less than $10$ events in onen year). See Fig.~\protect\ref{genius_b}.
}
\begin{tabular}{lll}
Experiment & Exp. constraint 
& Lower Bound on $M_N \,$ 
(GeV)\cr
Heidelberg-Moscow & $ T_{1/2} > 7.4 \times 10^{24} $ yr  & $
M_N > \sim 250$ \cr
GENIUS $1$ yr & $T_{1/2} > 6.0 \times 10^{27} $ yr 
& $M_N > \sim 700$ \cr
GENIUS $4$ yr & $T_{1/2} > 2.3 \times 10^{28} $ yr 
& $M_N > \sim  850$ \cr
LHC & $N_{events} < 10 $ & $M_N > \sim  850$\cr
\end{tabular}
\end{table}

\begin{figure}
\leavevmode
\caption{Parton level mechanism for production of Like-Sign-Di-leptons 
         (LSD) in high energy hadronic collisions. 
         The shaded blob contains all contributing diagrams for the virtual
         subprocess $W^+W^+ \to \ell^+\ell^+ $.   
         }
\label{fig1}
\end{figure}
\begin{figure}
\leavevmode
\caption{Production of LSD through quark-antiquark
         scattering. There are  here two interfering mechanisms to be 
         considered : 
         (a) virtual W fusion;
         (b) $\ell^+ N_{\ell} $  production 
         via  quark-antiquark annihilation 
         with subsequent hadronic decay of the heavy neutrino 
         $N_{\ell} \to \ell^+ q \bar{q}$.
         }
\label{fig2}
\end{figure}

\begin{figure}
\leavevmode
\caption{Comparison between the $\beta\beta_{0\nu}$       and the LEP II 
         upper bound on the quantity $|f|/(\sqrt{2}M_N)$ as a function of the
         heavy neutrino mass $M_N$, with the choice $\Lambda_{\hbox{c}}=M_N$. 
         {\it Regions above the curves are excluded}. 
         }
\label{delphi_com}
\end{figure}

\begin{figure}[htb]
  \begin{center}
    \leavevmode
    \caption{Cross section normalized to $|f|=1 $, {\it i.e.}
             $\sigma_1 = \sigma /|f|^4$ with the choice
             $\Lambda_{\hbox{c}} = 1 \hbox{ TeV}$.
             $(a)\, u u     \to d d     + \ell^+\ell^+$;
             $(b)\, u \bar{d} \to d \bar{u} + \ell^+\ell^+$;
             $(c)\, \bar{d} \bar{d} \to \bar{u} \bar{u} +\ell^+\ell^+$,
             $(d)$ the {\it solid} line is the sum of the 
	     contributions  from 
             Fig.~4a, 4b, 4c including factorizable corrections according to 
             Eq.~A1,~A2; the {\it dashed} 
             line is the process $u \bar{d} \to s \bar{c}+ \ell^+\ell^+$
             according to Eq. A3.
             Finally the {\it solid-diamond} line in (d) is the total 
             cross section,
             $\sigma_1$, including the sum of the sub-leading contributions 
             reported  in Fig.~\protect\ref{lhc_a2}.}
    \label{lhc_a}
  \end{center}
\end{figure} 

\begin{figure}[htb]
  \begin{center}
    \leavevmode
    \caption{Sub-leading processes: 
             the {\it solid} line is the sum of $u \bar{s}$ collisions 
             Eq.~A4 and A5; 
             the {\it long-dashed} line is the process $ u c \to d s 
             +\ell^+\ell^+ $; the {\it dashed} line is the sum 
             of $c\bar{s}$ collisions, Eqs.~A6 and A7;
             the dot-dashed line is the the sum of $c\bar{d}$ collisions, 
             Eqs.~A8 and A9, scaled by a factor of 10.
             Finally the {\it solid-diamond} line 
             is the total contribution to $\sigma_1$ of the above processes.}
    \label{lhc_a2}
  \end{center}
\end{figure} 

\begin{figure}[htb]
\begin{center}
    \leavevmode
    \caption{Same as in Fig.~\protect\ref{lhc_a} but  with the choice
             $\Lambda_{\hbox{c}} = M_N$. As explained in the text
             the different shape of the cross-section $\sigma_1$ 
             as function of $M_N$ respect to Fig.~\protect\ref{lhc_a}
             is because $\sigma_1 \propto \Lambda_{\hbox{c}}^{-4}$. Thus fixing
             $\Lambda_{\hbox{c}}=M_N$ gives of course a 
             different function of $M_N$ than choosing 
             $\Lambda_{\hbox{c}}= 1$ TeV. 
             The solid-diamond line in (d) again describes the total 
             $\sigma_1$ as done in Fig.~\protect\ref{lhc_a}. }
    \label{lhc_b}
\end{center}
\end{figure}   

\begin{figure}[htb]
  \begin{center}
    \leavevmode
    \caption{Sensitivity of LHC vs. current and next generation (GENIUS)
             double beta  experiments to 
             the compositeness parameters. Case $(a)$
             $\Lambda_{\hbox{c}} = 1 \, \hbox{TeV}$.
             {\it Non-observation of the signal excludes regions 
             above the curves}. If no signal will be observed 
             both LHC and GENIUS will be able to get
             upper bounds on $|f|$ stronger by almost an order of magnitude  
             respect to the present Heidelberg Moscow bound.
             There is a region where the LHC bound is weaker than the
             GENIUS 1 yr (4 yr) bound $M_N < 550\, (1000)$ GeV while 
             for $M_N > 550\, (1000)$ GeV
             the LHC bound is stronger.
             }
    \label{genius_a}
  \end{center}
\end{figure}  

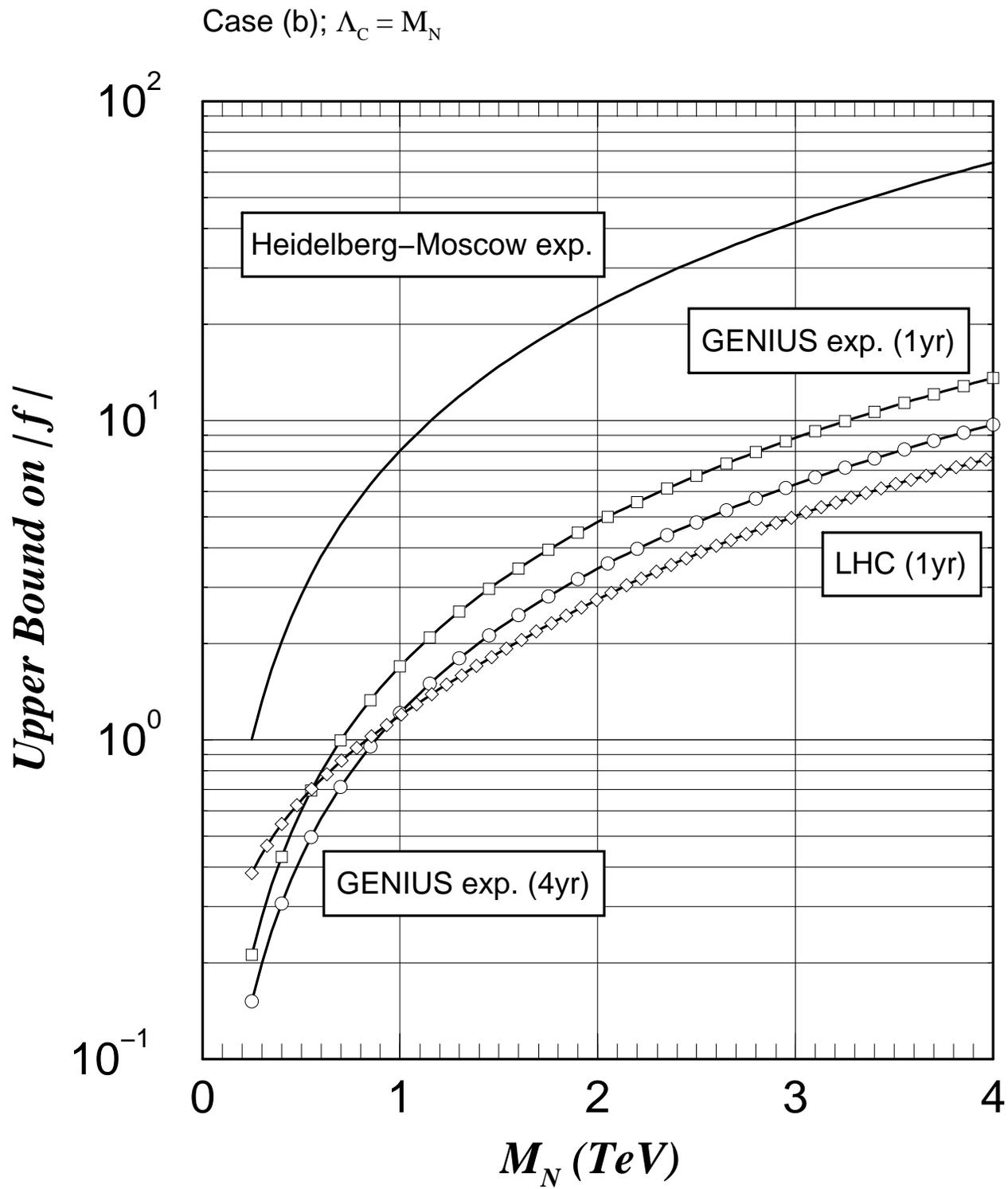
\begin{figure}[htb]
  \begin{center}
    \leavevmode
    \caption{Same as in Fig~\protect\ref{genius_a} but with
             $\Lambda_{\hbox{c}} = M_N $. Also here 
             {\it regions above the curves are excluded}.
             Here the shape of the LHC exclusion plot is similar to that
             of $\beta\beta_{0\nu}$. The values of $M_N$ at which the LHC curve
             crosses those of GENIUS are the same as  
             in Fig.~\protect\ref{genius_a}. 
             }
    \label{genius_b}
  \end{center}
\end{figure}  


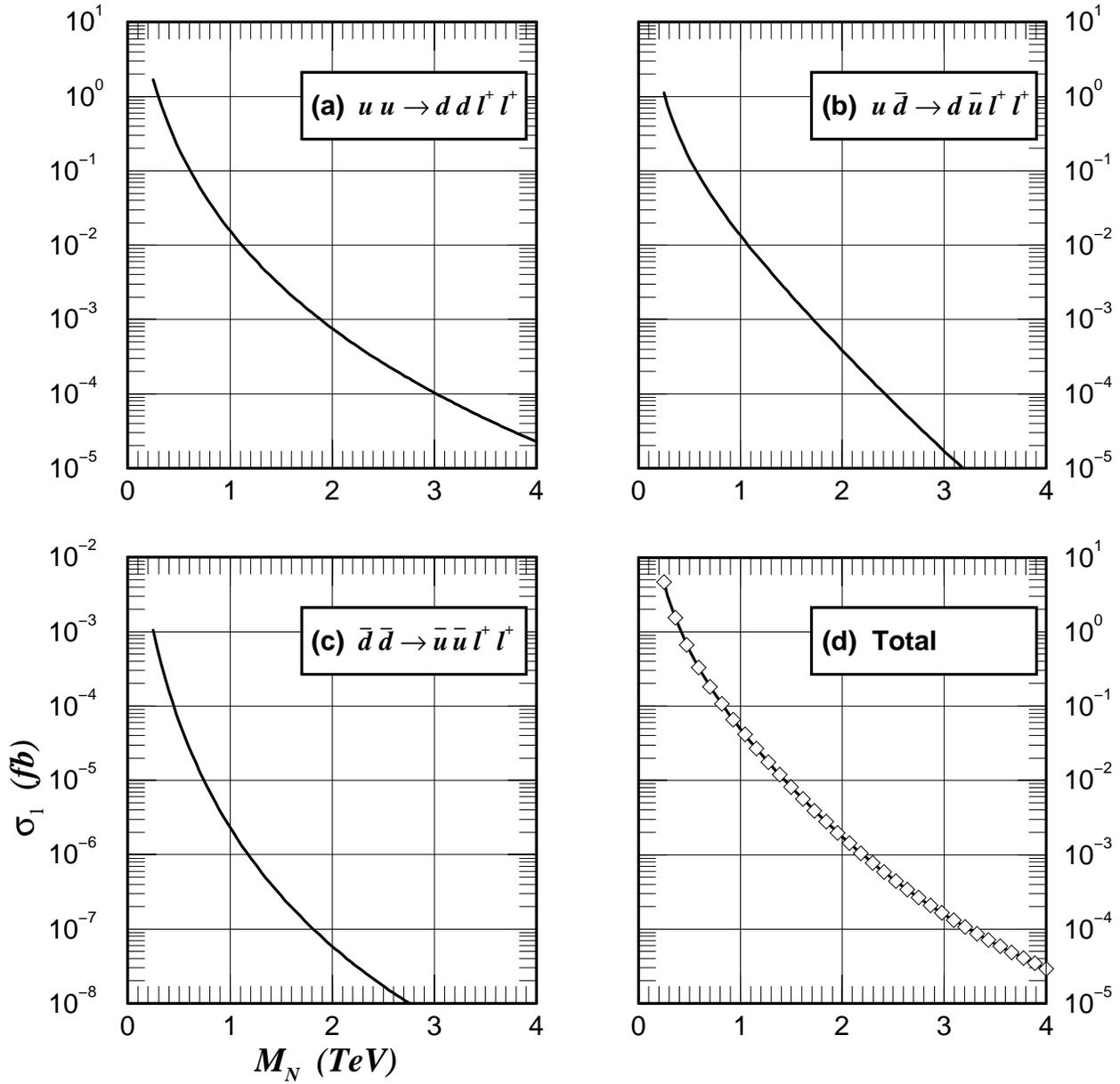
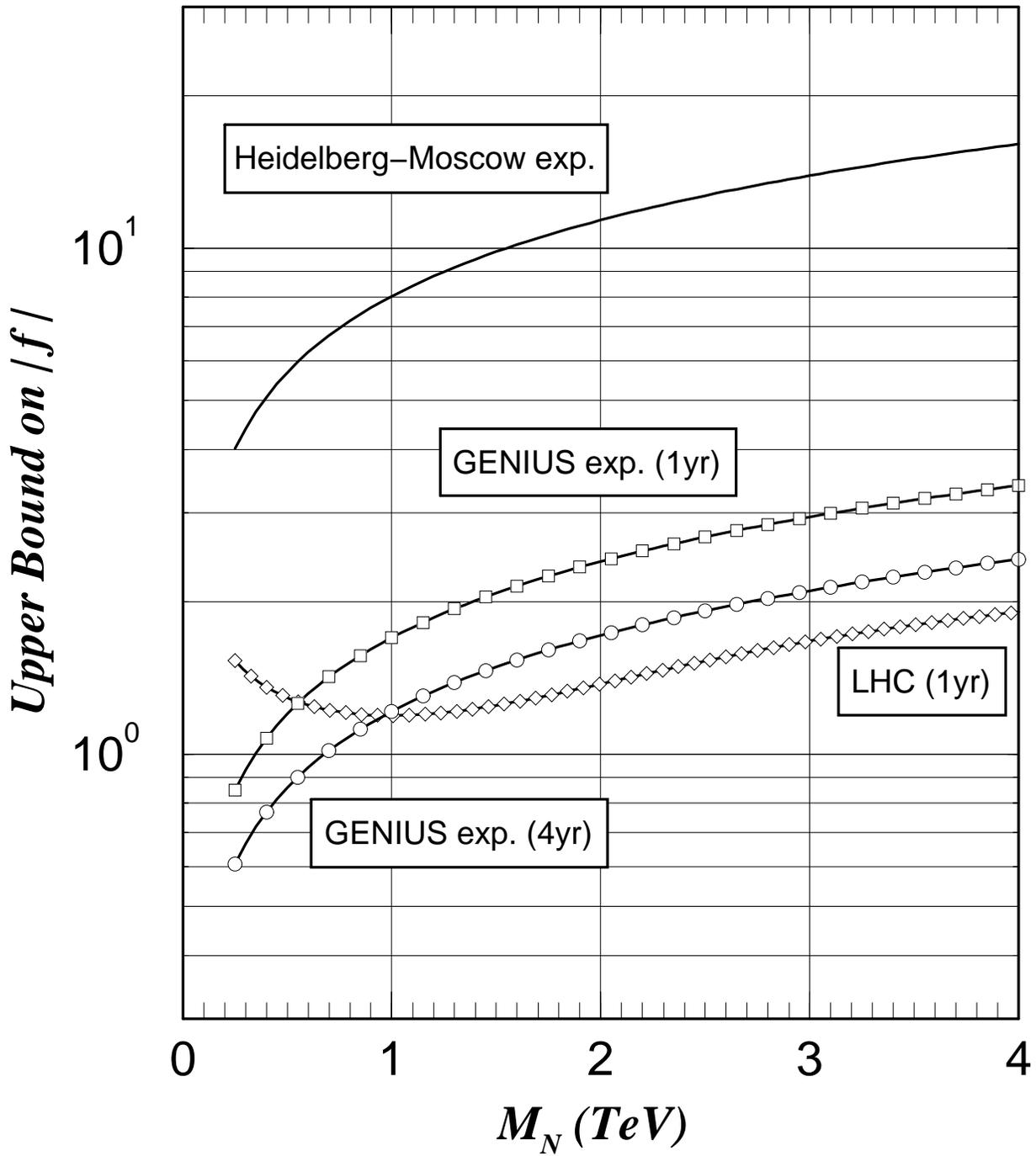
\end{document}